\documentclass[showemail,pre]{revtex4}

\usepackage{graphicx}   
\usepackage{bm}
\begin{document}

\title{
A Biological Coevolution Model with Correlated Individual-Based Dynamics
}
\author{Volkan Sevim}\email{sevim@csit.fsu.edu}
\author{Per Arne Rikvold}\email{rikvold@csit.fsu.edu}
\affiliation{
School of Computational Science and Information Technology,\\
Center for Materials Research and Technology, and Department of Physics,\\
Florida State University, Tallahassee, FL 32306-4120, USA\\
}
\date{\today}

\begin{abstract}
We study the effects of \index{interspecific correlations}interspecific correlations in a \index{biological coevolution}biological coevolution model in which organisms are represented by genomes of bitstrings. We present preliminary results for this model, indicating that these correlations do not significantly affect the statistical behavior of the system.
\end{abstract}

\maketitle

\section{Introduction}
\label{sec:I}
The dynamics of biological coevolution poses many problems of interest to the statistical-physics and complex-systems communities \cite{DROS02, DROS01}. Recently, we studied a coevolution model that is a simplified version of the one introduced by Hall, {\it et al}. \cite{HALL02,CHRI02, RIKV03,RIKV32}. Our model, in which individuals give birth, mutate and die, displays \index{punctuated equilibria}punctuated equilibria-like, quiet periods interrupted by bursts of mass extinctions \cite{GOUL77,GOUL93}. Interactions in this model were given by a random interaction matrix. Here we report on a modified version of the model, in which we have added correlations to the \index{interaction matrix} interaction matrix in order to increase the biological realism. The modified model is compared with the original one to assess the effects of correlations.

\section{Model}
\label{sec:M}
We used a \index{bitstring}bitstring genome of length $L$ to model organisms in the \index{Monte Carlo}Monte Carlo (MC) simulations \cite{EIGE71, EIGE88}. In this sense, each genotype, which is just an $L$-bit number, is considered a different haploid species. Therefore, the terms ``genotype" and ``species" are used in the same sense in this paper. We denote the population of species $I$ at a discrete time $t$ as $n_I(t)$, and several of the $2^L$ possible species can be present at the same time in our ``ecosystem." All species reproduce asexually (by cloning) in discrete, non-overlapping generations. In each generation $t$, every individual of species $I$ is allowed to give birth to $F$ offspring with a probability $P_I$. Whether it reproduces or not, the individual dies at the end of the generation, so that only offspring can survive to the next generation. The reproduction probability for an individual of species $I$ is given by \cite{ HALL02,CHRI02,RIKV03,RIKV32}

\begin{equation}
P_I(\{n_J(t)\})
=
\frac{1}{1 + \exp\left[ - \sum_J M_{IJ} n_J(t)/N_{\rm tot}(t)
+ N_{\rm tot}(t)/N_0 \right]}
\;.
\label{eq:P}
\end{equation}
The Verhulst factor $N_0$ represents the carrying capacity of the ``ecosystem" and prevents the total population $N_{\rm tot}(t) = \sum_I n_I(t)$ from diverging to infinity \cite{VERH1838}.  $\bf M$ is the interaction matrix, in which a matrix element $M_{IJ}$ represents the effect of the population density of species $J$ on species $I$. A positive $M_{IJ}$ means that species $I$ benefits from species $J$, while a negative $M_{IJ}$ corresponds to a situation in which species $I$ is harmed or inhibited by the presence of $J$. The form of the interaction matrix  $\bf M$ is discussed in the next section. 

In each generation, all individuals undergo mutation with a probability $\mu$. If an individual is chosen to mutate, one bit in its genome is picked randomly and flipped. Since we consider different genotypes as different species, mutations lead to speciation, i.e., creation of another species.

\section{The Interaction Matrix}
\label{sec:S}
In this study, the interaction matrix  $\bf M$ was set up in four different ways. In the first case, all off-diagonal elements were randomly and uniformly distributed on $[-1, 1]$, while all diagonal elements were set to zero. This corresponds to the case in Refs. \cite{RIKV03, RIKV32}. We shall call this the uniform-uncorrelated model.
In the second case, we added correlations between interaction constants of similar species to make this model more realistic. In a real ecosystem, two different but closely related species $X$ and $Y$ interact with another species $Z$ in a similar way. Therefore, the interaction constant $M_{XZ}$ should be positively correlated with $M_{YZ}$. To implement these correlations, we modified the interaction matrix by averaging all terms over their nearest neighbors (nn) in Hamming space. Thus, 

\begin{figure}[t]
\vspace{0.4truecm}
\begin{center}
\includegraphics[width=.455\textwidth]{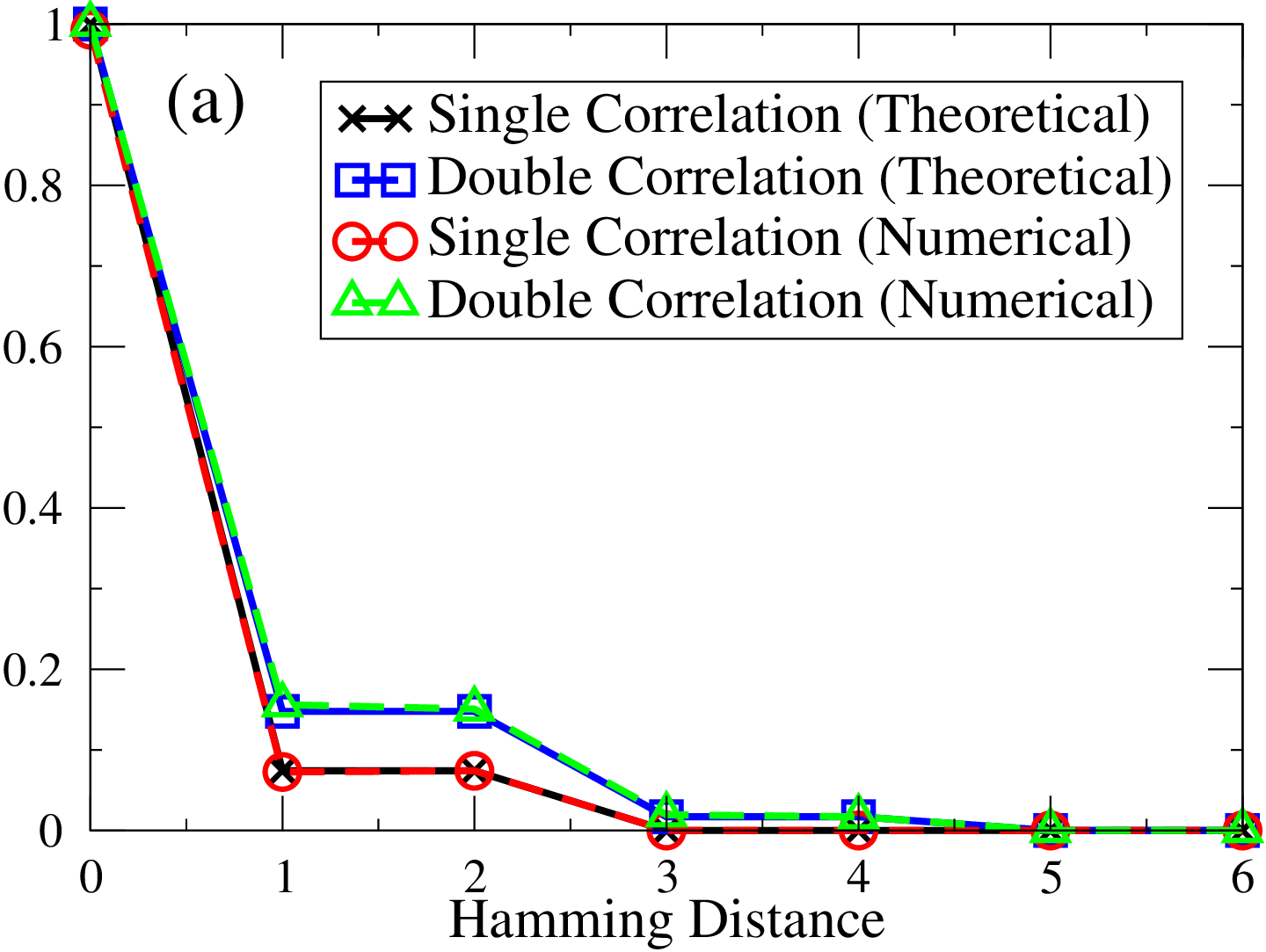}
\hspace{0.35truecm}
\includegraphics[width=.445\textwidth]{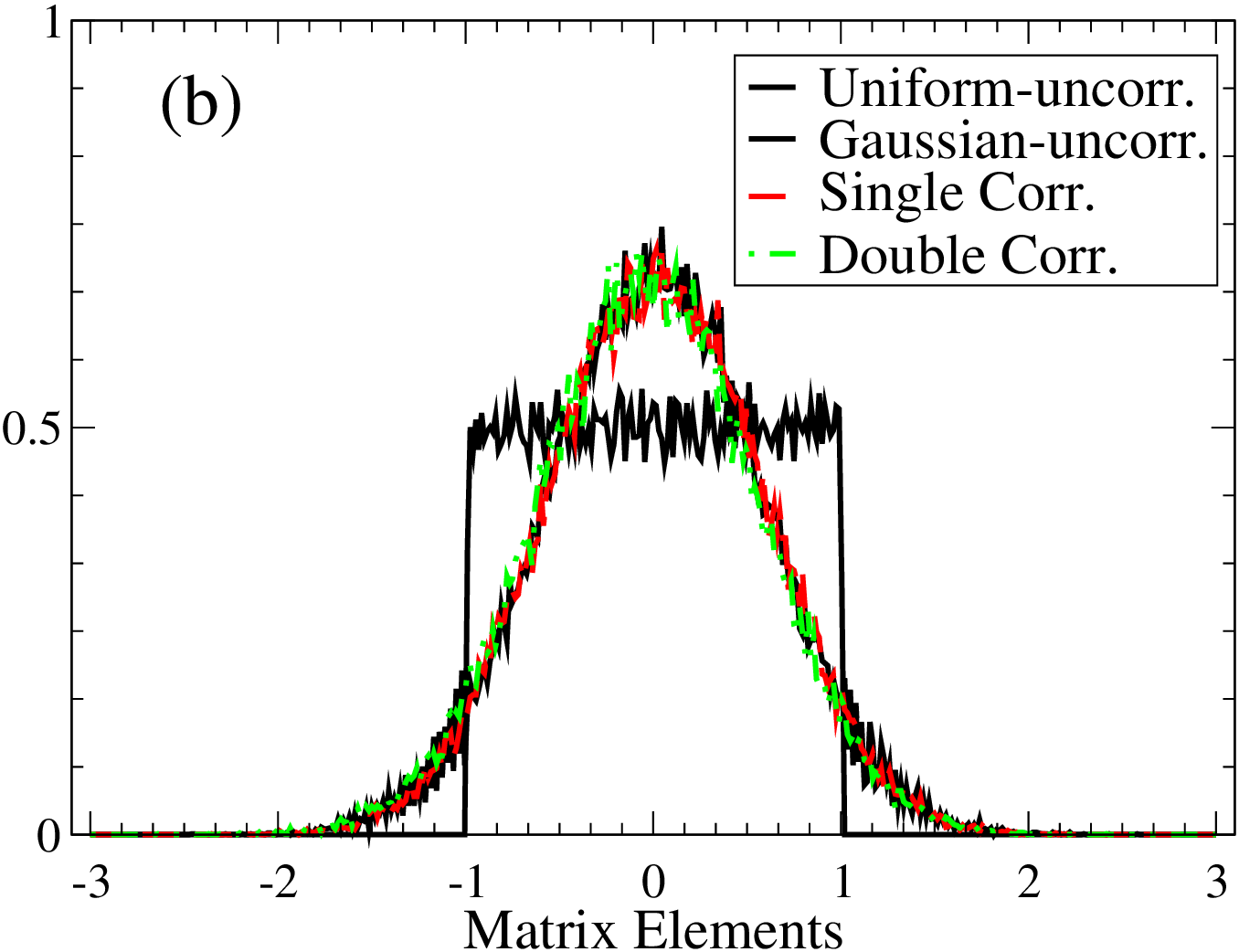}
\end{center}
\caption[]{
Correlation functions and matrix-element distributions after averaging. 
 ({\bf a})
Theoretical (solid) and numerical (dashed) results for the correlation functions for $2^{13} \times 2^{13}$ interaction matrices of the single-correlated (lower) and double-correlated (upper) models. The theoretical and numerical results agree.
 ({\bf b})
Distributions of matrix elements: uniform and Gaussian uncorrelated (solid curves with square and Gaussian distributions, respectively), single-correlated (dashed curve) and double-correlated (dot-dashed curve) models with $L=8$. The approximately Gaussian distributions for the two correlated models and the distribution for the Gaussian-uncorrelated model practically overlap
}
\label{fig:fig1}
\end{figure}

\begin{equation}
M_{IJ}= \left[ M^0_{IJ}+ \sum_{(K,L)\in nn(I,J)} M^0_{KL}  \right]/ (2L+1)^{1/2} \;,
\label{eq:NNAV}
\end{equation}
where $M^0_{IJ}$ are independent variables uniformly distributed on $[-1,1]$, and $nn(I,J)$ are those bitstring pairs that differ from the bitstring pair $(I,J)$ by one bit (a Hamming distance of one). A square-root appears in the denominator because we multiply the average with the square-root of the normalization constant in order not to change the standard deviation of the matrix elements. 

To investigate the results of longer-range correlations we also modified the random interaction matrix by averaging all terms over their nearest and next-nearest neighbors (nnn) in Hamming space:

\begin{equation}
M_{IJ}= \left[ M^0_{IJ}+ \sum_{(K,L)\in nn(I,J)} M^0_{KL} + \sum_{(K,L)\in nnn(I,J)} M^0_{KL}\right]/ (2L^2+L+1)^{1/2}
\label{eq:NNNAV}
\end{equation}
where $nnn(I,J)$ is the set of bitstring pairs that differ from $(I,J)$ by two bits. We shall call the nn-averaged and nnn-averaged models single-correlated and double-correlated, respectively.

\begin{figure}[t]
\centering
\vspace{0.74truecm}
\includegraphics[width=.75\textwidth]{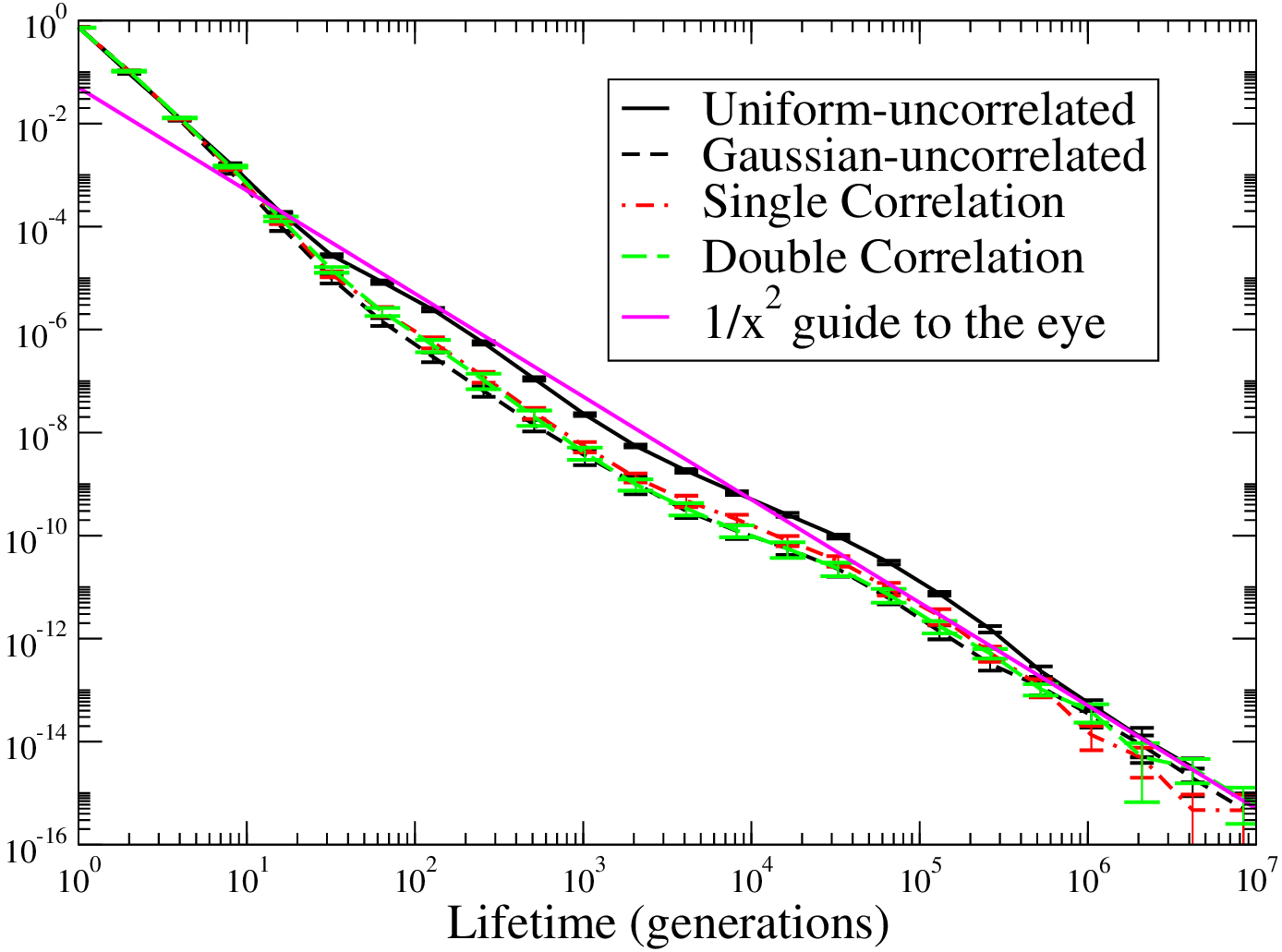}
\caption[]{Normalized histograms of species-lifetimes based on simulations of $2^{25}$ generations each: uniform-uncorrelated (solid curve), Gaussian-uncorrelated (short-dashed curve), single-correlated (dot-dashed curve) and double correlated (long-dashed curve). The distributions for the models with a Gaussian matrix-element distribution overlap (within the margin of error), and they differ significantly from the lifetime distribution for the uniform-uncorrelated model. The histograms exhibit a power-law like decay with an exponent near $-2$. Results are averaged over eight runs each
}
\label{fig:fig2}
\end{figure}

As a result of the central limit theorem, after averaging, the distributions of the matrix elements in both the single and double-correlated models take an approximately Gaussian form with the same standard deviation as the uniform distribution of the elements of the initial, random matrix (see Fig. \ref{fig:fig1}b). In order to see whether a possible difference in the results for different models is due to the correlations or the matrix-element distributions, we also set up another uncorrelated interaction matrix. In this fourth model, the uncorrelated, random matrix elements are distributed with a Gaussian distribution with the same standard deviation, $\sqrt{1/3}$, as in the correlated models. We shall call this the Gaussian-uncorrelated model.

The correlation functions and the distributions of interaction-matrix elements for all the models are shown in Fig.~\ref{fig:fig1}. The steps in the correlation functions are a result of the chosen metric. We use a city-block metric in which the Hamming distance between two matrix elements $M_{IJ}$ and $M_{KL}$ is given by $H(I,K)+H(J,L)$ where $H(I,K)$ is the Hamming distance between two $L$-bit bitstrings, $I$ and $K$.

\section{Simulation Results}
\label{sec:C}

The interaction matrix is set up at the beginning and is not modified during the course of the simulation. For all models, we performed eight sets of simulations for $2^{25}$=33\,554\,432 generations with the same parameters as in Refs. \cite{RIKV03, RIKV32}: genome length $L=13$,
mutation rate $\mu = 10^{-3}$ per individual per generation,
carrying capacity $N_0 = 2000$, and fecundity $F=4$. We began the simulations with 200 individuals of genotype $0$. 
In order to compare the models, we constructed histograms corresponding to the species-lifetime distributions. The lifetime of a species is defined as the number of generations between its creation and extinction. As seen in Fig.\ref{fig:fig2}, the \index{lifetime distribution}lifetime distributions of the correlated and Gaussian-uncorrelated models overlap within the margin of error, and they differ significantly from the lifetime distribution of the uniform-uncorrelated model. They all exhibit a power-law like decay with an exponent near $-2$.  The correlations between the matrix elements do not seem to affect the behavior of the lifetime distribution to a statistically significant degree, at least not for the relatively weak correlations that were introduced here. On the other hand, changes in the marginal probability density of the individual matrix elements do have statistically significant effects, even though gross features, such as the approximate $1/x^{2}$ behavior of the species-lifetime distribution are not changed. Similar conclusions are reached also for other quantities that we studied. In particular, the power-spectral density of the Shannon-Wiener species diversity index shows \index{1/f noise}$1/f$ noise \cite{RIKV03,RIKV32} with an overall intensity that depends more on the marginal matrix-element distribution than on correlations in  $\bf M$.

Although we have tested only weak correlations, and so our conclusion is only preliminary, there appears to be no disadvantage in using a random interaction matrix to model such an ``ecosystem." This has obvious computational advantages as it makes it possible to simulate systems with larger numbers of completely different species.

\section*{Acknowledgments}
This research was supported by U.S.\ National Science Foundation Grant
No. DMR-0240078, and by Florida State University through the School of Computational
Science and Information Technology and the Center for Materials Research
and Technology.

%
%

\begin{thebibliography}{10}

\bibitem{DROS02}
B.~Drossel, A.J. McKane, in: {\it Handbook of Graphs and Networks: From the Genome to the Internet}, Ed. by S. Bornholdt, H. G. Schuster (Wiley-VCH, Berlin 2002)

\bibitem{DROS01}
B.~Drossel: Adv.\ Phys.\ {\bf 50},  209  (2001)

\bibitem{HALL02}
M.~Hall, K.~Christensen, S.A.~di~Collobiano, H.J.~Jensen: 
Phys.\ Rev.\ E  {\bf 66},  011904  (2002)

\bibitem{CHRI02}
K.~Christensen, S.A.~di~Collobiano, M.~Hall, H.J.~Jensen: 
J.\ Theor.\  Biol.\ {\bf 216},  73  (2002)

\bibitem{RIKV03}
P.A.~Rikvold, R.K.P.~Zia: Phys.\ Rev.\ E {\bf 68}, 031913 (2003), and references therein

\bibitem{RIKV32}
P.A.~Rikvold, R.K.P.~Zia, in: {\it Computer Simulation Studies in Condensed-Matter Physics XVI.} Ed. by D. P. Landau, S. P. Lewis, H.-B. Sch\"uttler (Springer-Verlag, Berlin Heidelberg New York, 2004)

\bibitem{GOUL77}
S.J.~Gould, N.~Eldredge: Paleobiology {\bf 3},  115  (1977)

\bibitem{GOUL93}
S.J.~Gould, N.~Eldredge: Nature {\bf 366},  223  (1993)


\bibitem{EIGE71}
M.~Eigen: Naturwissenschaften {\bf 58},  465  (1971)

\bibitem{EIGE88}
M.~Eigen, J.~McCaskill, P.~Schuster: 
J.\ Phys.\ Chem.\ {\bf 92},  6881  (1988)

\bibitem{VERH1838}
P.F.~Verhulst: Corres.\ Math.\ et Physique {\bf 10},  113  (1838)

 
\end{thebibliography}
%

\end{document}